\documentclass[twoside,slac_one]{revtex4}
\usepackage{graphicx}
\usepackage{fancyhdr}
\usepackage{amsmath} 
\usepackage{bm}
\usepackage{amsxtra}
\usepackage{amssymb}
\usepackage{amsthm}
\usepackage{latexsym}
\usepackage{lscape}

\input babarsym

\usepackage{relsize}
\def\babar{\mbox{\slshape B\kern-0.1em{\smaller A}\kern-0.1em
    B\kern-0.1em{\smaller A\kern-0.2em R}}}

\def\en         {\ensuremath{e^-}\xspace}   
\def\ep         {\ensuremath{e^+}\xspace}
 
\def\epem       {\ensuremath{e^+e^-}\xspace}

\def\mup        {\ensuremath{\mu^+}\xspace}
\def\mun        {\ensuremath{\mu^-}\xspace} 
\def\mumu       {\ensuremath{\mu^+\mu^-}\xspace}

\def\taup       {\ensuremath{\tau^+}\xspace}
\def\taum       {\ensuremath{\tau^-}\xspace}

\def\ellell     {\ensuremath{\ell^+ \ell^-}\xspace}

\def\nub        {\ensuremath{\overline{\nu}}\xspace}
\def\nunub      {\ensuremath{\nu{\overline{\nu}}}\xspace}
\def\nub        {\ensuremath{\overline{\nu}}\xspace}
\def\nunub      {\ensuremath{\nu{\overline{\nu}}}\xspace}
\def\nue        {\ensuremath{\nu_e}\xspace}
\def\nueb       {\ensuremath{\nub_e}\xspace}

\def\num        {\ensuremath{\nu_\mu}\xspace}
\def\numb       {\ensuremath{\nub_\mu}\xspace}

\def\nut        {\ensuremath{\nu_\tau}\xspace}
\def\nutb       {\ensuremath{\nub_\tau}\xspace}

\def\nul        {\ensuremath{\nu_\ell}\xspace}
\def\nulb       {\ensuremath{\nub_\ell}\xspace}

\def\g     {\ensuremath{\gamma}\xspace}
\def\gaga  {\ensuremath{\gamma\gamma}\xspace}  
\def\ggstar{\ensuremath{\gamma\gamma^*}\xspace}

\def\qqbar {\ensuremath{q\overline q}\xspace}

\def\t     {\ensuremath{t}\xspace}
\def\tbar  {\ensuremath{\overline t}\xspace}
\def\tbar  {\ensuremath{\overline t}\xspace}

\def \tb     {\ensuremath{\tbar}\xspace}
\def \ttb    {\ensuremath{\t {\kern -0.16em \tb}}\xspace}

\def\piz   {\ensuremath{\pi^0}\xspace}

\def\ppz   {\ensuremath{\pi^0\pi^0}\xspace}
\def\pip   {\ensuremath{\pi^+}\xspace}
\def\pim   {\ensuremath{\pi^-}\xspace}
\def\pipi  {\ensuremath{\pi^+\pi^-}\xspace}

\def\pimp  {\ensuremath{\pi^\mp}\xspace}

\def\kaon  {\ensuremath{K}\xspace}
\def\Kbar  {\kern 0.2em\overline{\kern -0.2em K}{}\xspace}

\def\Kz    {\ensuremath{K^0}\xspace}
\def\Kzb   {\ensuremath{\Kbar^0}\xspace}
\def\KzKzb {\ensuremath{\Kz \kern -0.16em \Kzb}\xspace}
\def\Kp    {\ensuremath{K^+}\xspace}
\def\Km    {\ensuremath{K^-}\xspace}
\def\Kpm   {\ensuremath{K^\pm}\xspace}

\def\KpKm  {\ensuremath{\Kp \kern -0.16em \Km}\xspace}
\def\KS    {\ensuremath{K^0_{\scriptscriptstyle S}}\xspace} 
\def\KL    {\ensuremath{K^0_{\scriptscriptstyle L}}\xspace} 
\def\Kstarz  {\ensuremath{K^{*0}}\xspace}

\def\Kstar   {\ensuremath{K^*}\xspace}

\def\Kstarp  {\ensuremath{K^{*+}}\xspace}

\def\Dbar    {\kern 0.2em\overline{\kern -0.2em D}{}\xspace}

\def\Dz      {\ensuremath{D^0}\xspace}
\def\Dzb     {\ensuremath{\Dbar^0}\xspace}
\def\DzDzb   {\ensuremath{\Dz {\kern -0.16em \Dzb}}\xspace}
\def\Dp      {\ensuremath{D^+}\xspace}
\def\Dm      {\ensuremath{D^-}\xspace}

\def\DpDm    {\ensuremath{\Dp {\kern -0.16em \Dm}}\xspace}
\def\Dstar   {\ensuremath{D^*}\xspace}

\def\Dstarz  {\ensuremath{D^{*0}}\xspace}

\def\Dstarp  {\ensuremath{D^{*+}}\xspace}
\def\Dstarm  {\ensuremath{D^{*-}}\xspace}

\def\B       {\ensuremath{B}\xspace}
\def\Bpr     {\ensuremath{B^{*}}\xspace}
\def\Bs      {\ensuremath{B_{s}}\xspace}
\def\Bss     {\ensuremath{B_{s}^{*}}\xspace}
\def\Bsss    {\ensuremath{B_{s}^{(*)}}\xspace}

\def\Bbar    {\kern 0.18em\overline{\kern -0.18em B}{}\xspace}
\def\Bprb    {\kern 0.18em\overline{\kern -0.18em B^{*}}{}\xspace}
\def\BBpr    {\ensuremath{\B {\kern -0.16em \Bpr}}\xspace}
\def\BBprb   {\ensuremath{\B {\kern -0.16em \Bprb}}\xspace}
\def\BprBprb {\ensuremath{\Bpr {\kern -0.16em \Bprb}}\xspace}
\def\Bsbar   {\kern 0.18em\overline{\kern -0.18em B_{s}}{}\xspace}
\def\Bssbar  {\kern 0.18em\overline{\kern -0.18em B_{s}^{*}}{}\xspace}
\def\Bsssbar {\kern 0.18em\overline{\kern -0.18em B_{s}^{(*)}}{}\xspace}
\def\Bb      {\ensuremath{\Bbar}\xspace}

\def\BB      {\ensuremath{\B {\kern -0.16em \Bb}}\xspace}
\def\BBp     {\ensuremath{\B {\kern -0.16em \Bb}\pi}\xspace}
\def\BBpp    {\ensuremath{\B {\kern -0.16em \Bb}\pi\pi}\xspace}
\def\BBprp   {\ensuremath{\B {\kern -0.16em \Bprb}\pi}\xspace}
\def\BsBsb   {\ensuremath{\Bs {\kern -0.16em \Bsbar}}\xspace}
\def\BsBssb  {\ensuremath{\Bs {\kern -0.16em \Bssbar}}\xspace}
\def\BssbBs  {\ensuremath{\Bssbar {\kern -0.16em \Bs}}\xspace}
\def\BssBsb  {\ensuremath{\Bss {\kern -0.16em \Bsbar}}\xspace}
\def\BssBssb {\ensuremath{\Bss {\kern -0.16em \Bssbar}}\xspace}
\def\Bz      {\ensuremath{B^0}\xspace}
\def\Bsz     {\ensuremath{B_{s}^0}\xspace}
\def\Bzb     {\ensuremath{\Bbar^0}\xspace}
\def\Bszb    {\ensuremath{\Bsbar^0}\xspace}
\def\BzBzb   {\ensuremath{\Bz {\kern -0.16em \Bzb}}\xspace}
\def\BszBszb {\ensuremath{\Bsz {\kern -0.16em \Bszb}}\xspace}
\def\BsssBsssb {\ensuremath{\Bsss {\kern -0.16em \Bsssbar}}\xspace}

\def\Bu      {\ensuremath{B^+}\xspace}
\def\Bub     {\ensuremath{B^-}\xspace}

\def\Bpm     {\ensuremath{B^\pm}\xspace}

\def\BpBm    {\ensuremath{\Bu {\kern -0.16em \Bub}}\xspace}
\def\Bs      {\ensuremath{B_s}\xspace}

\def\BorBbar    {\kern 0.18em\optbar{\kern -0.18em B}{}\xspace}
\def\DorDbar    {\kern 0.18em\optbar{\kern -0.18em D}{}\xspace}
\def\KorKbar    {\kern 0.18em\optbar{\kern -0.18em K}{}\xspace}

\def\jpsi     {\ensuremath{{J\mskip -3mu/\mskip -2mu\psi\mskip 2mu}}\xspace}

\mathchardef\Upsilon="7107
\def\Y#1S{\ensuremath{\Upsilon{(#1S)}}\xspace}

\def\FourS {\Y4S}

\mathchardef\Deltares="7101
\mathchardef\Xi="7104
\mathchardef\Lambda="7103
\mathchardef\Sigma="7106
\mathchardef\Omega="710A

\def\Deltabar{\kern 0.25em\overline{\kern -0.25em \Deltares}{}\xspace}
\def\Lbar{\kern 0.2em\overline{\kern -0.2em\Lambda\kern 0.05em}\kern-0.05em{}\xspace}
\def\Sigbar{\kern 0.2em\overline{\kern -0.2em \Sigma}{}\xspace}
\def\Xibar{\kern 0.2em\overline{\kern -0.2em \Xi}{}\xspace}
\def\Obar{\kern 0.2em\overline{\kern -0.2em \Omega}{}\xspace}
\def\Nbar{\kern 0.2em\overline{\kern -0.2em N}{}\xspace}
\def\Xb{\kern 0.2em\overline{\kern -0.2em X}{}\xspace}

\def\X {\ensuremath{X}\xspace}

\def\BR         {{\ensuremath{\cal B}\xspace}}

\def\bpsikl     {\ensuremath{\Bz \to \jpsi \KL}\xspace}

\def\Bzbtox     {\ensuremath{\Bzb \to \X}\xspace}

\def\mus  {\ensuremath{\rm \,\mus}\xspace}

\def\mus        {\ensuremath{\,\mu{\rm s}}\xspace}    

\def\degrees{\ensuremath{^{\circ}}\xspace}

\def\to                 {\ensuremath{\rightarrow}\xspace}

\def\pep2{PEP-II}
\def\BF{$B$ Factory}

\def\gsim{{~\raise.15em\hbox{$>$}\kern-.85em
          \lower.35em\hbox{$\sim$}~}\xspace}
\def\lsim{{~\raise.15em\hbox{$<$}\kern-.85em
          \lower.35em\hbox{$\sim$}~}\xspace}

\def\epstag  {\ensuremath{\varepsilon_{\rm tag}}\xspace}

\def\jetset74   {\mbox{\tt Jetset \hspace{-0.5em}7.\hspace{-0.2em}4}\xspace}

\usepackage{relsize}

\begin{document}


\title{\babar\ Results on CP Violation in B Decays}

%


\author{Romulus Godang (On Behalf of the \babar\ Collaboration)}
\affiliation{Department of Physics, University of South Alabama, Mobile, Alabama 36688}

\begin{abstract}
We report on the study of the decay $B^+ \to D^0(\bar{D0}) K^+$ where $D^0$ and $\bar{D0}$ decaying
to $K \pi \pi^0$, with the Atwood Dunietz and Soni (ADS) method. We measure the ratios Rads,
$R^+$, $R^-$ since the processes $B^+ \to D^0 \bar{K^+}$ and $B^+ \to D^0 K^+$ are proportional 
to $V_{cb}$ and $V_{ub}$, respectively, are sensitive to $r_B$ and to the weak phase $\gamma$.
\end{abstract}

\maketitle

\thispagestyle{fancy}


\section{Introduction}

\begin{figure}[h]
\begin{center}
\begin{tabular}{lr}\hspace{-4mm}
\includegraphics[width=5.2cm]{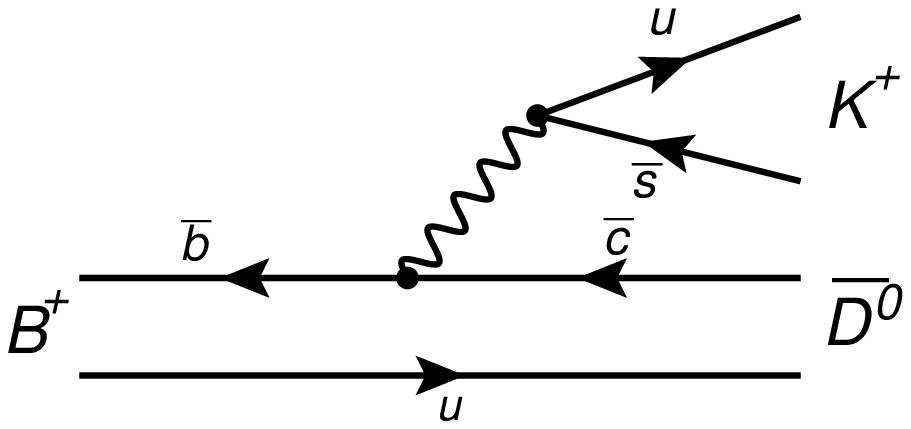}& \hspace{2cm}  
\includegraphics[width=5.2cm]{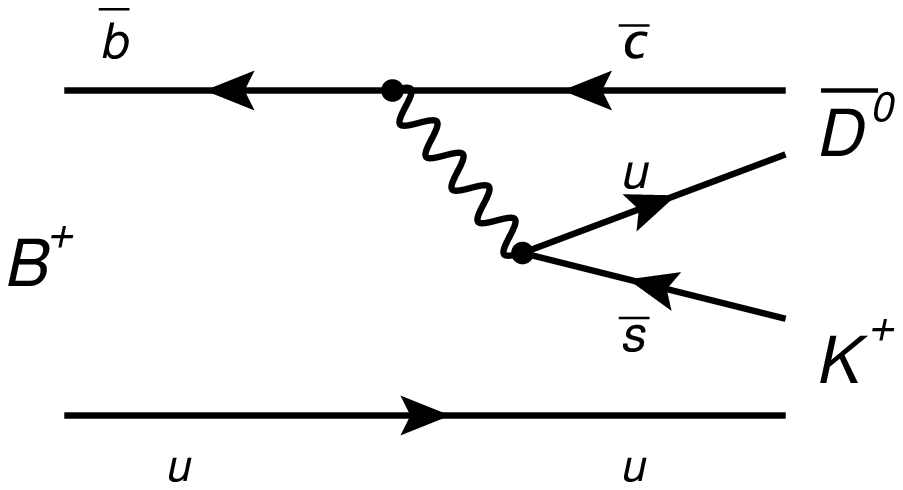} 
\end{tabular}
\caption{Feynman diagrams for $B^+ \to \bar{D}^0 K^+$ ($b \to u \bar{c} s$))}
\label{bc_diagram}
\end{center}
\end{figure}

\begin{figure}[h]
\begin{center}
\begin{tabular}{lr}\hspace{-4mm}
\includegraphics[width=5.2cm]{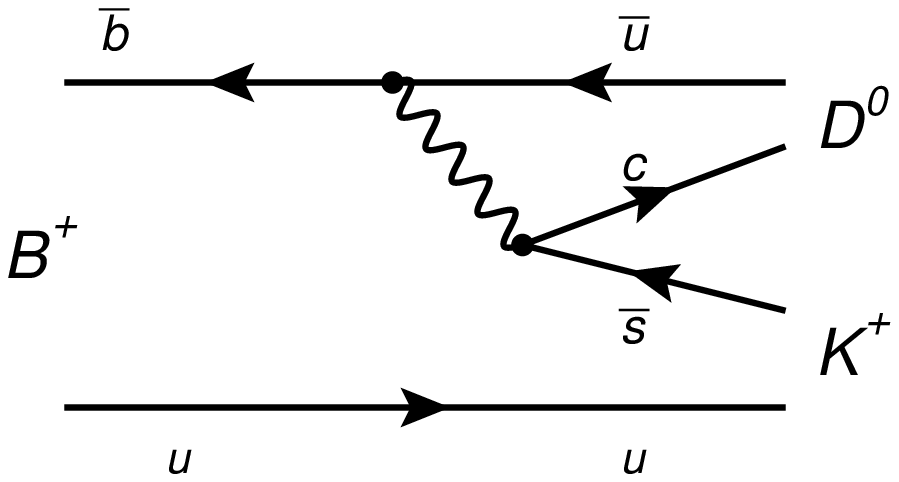}& \hspace{2cm}  
\includegraphics[width=5.2cm]{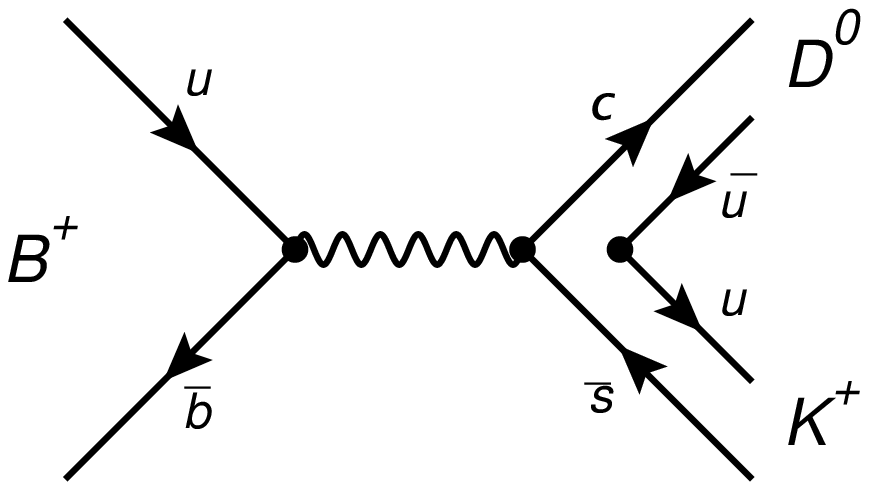} 
\end{tabular}
\caption{Feynman diagrams for $B^+ \to D^0 K^+$ ($b \to c \bar{u} s$)}
\label{bu_diagram}
\end{center}
\end{figure}

During recent years, several methods have been proposed to obtain the information
on the Cabibbo-Kobayashi-Maskawa (CKM)~\cite{cabibbo,KM}, phase angle $\gamma$. 
In the Standard Model (SM), the angle $\gamma$ is the relative phase between 
$b \to c \bar{u} s$ and $b \to u \bar{c} s$ transitions as indicated in 
Fig.~\ref{bc_diagram} and Fig.~\ref{bu_diagram}.
 
This phase angle $\gamma$ can be measured using a variety of methods involving 
$B$-meson decays that mediated by either only tree-level or both tree-level 
and loop-level amplitudes. A theoretical source of information on the angle $\gamma$ is 
provided by $B^- \to D^{(*)} K^-$ decays. The $ D^{(*)}$ represents an admixture of 
$D^{(*)0}$ and $\bar{D}^{(*)0}$ states. These decays exploit the interference between
$B^- \to D^{(*)0} K^-$ and $B^- \to \bar{D}^{(*)0} K^-$ that occurs when the 
$D^{(*)0}$ and $\bar{D}^{(*)0}$ decay to common final states.

In the Gronau-London-Wyler (GLW) method~\cite{GLW1,GLW2}, the $D^0$ meson is reconstructed based on 
Cabibbo-suppressed to $CP$-eigenstates, such as $K^+ K^-$. In order to determine the angle $\gamma$
from $B^{\pm}$ decays, we define the two experimental observables direct-$CP$-violating partial 
decay rate asymmetries as 
\begin{equation}
{A_{CP\pm}} \equiv \frac{\Gamma(B^- \to D_{CP\pm} K^-)~-~ \Gamma(B^+ \to D_{CP\pm} K^+)}
                           {\Gamma(B^- \to D_{CP\pm} K^-)~+~\Gamma(B^+ \to D_{CP\pm} K^+)}
= \frac{\pm 2r_B sin \delta_B sin \gamma}{1 + r_B^2 \pm 2r_B cos \delta_B cos \gamma}
\label{eq-A_CP}
\end{equation}
We also define the two ratios of charged averaged partial rates using $D$ meson decays to
$CP$ and flavor eigenstates as
\begin{equation}
{R_{CP\pm}} \equiv 2 \frac{\Gamma(B^- \to D_{CP\pm} K^-) ~+~ 
             \Gamma(B^+ \to D_{CP\pm} K^+)} {\Gamma(B^- \to D^0 K^-)~+~\Gamma(B^+ \to \bar{D^0} K^+)}
= 1 + r_B^2 \pm 2r_B cos\delta_B cos\gamma
\label{eq-R_CP}
\end{equation}
where $D_{CP\pm}$ refer to the $CP$ eigenstates of $D$ meson system and $\delta_B$ is the difference 
of their strong phases, and $r_B$ is the magnitude of the ratio of the amplitudes for each decay
\begin{equation}
r_B \equiv \frac{|A(B^- \to \bar{D^0} K^-)|}{|A(B^- \to D^0K^-)|}
\label{eq-rB}
\end{equation}
In the Adwood-Dunietz-Soni (ADS) method~\cite{ADS1,ADS2}, the $D^0$ meson is reconstructed in 
the doubly Cabibbo-suppressed decay $D^0 \to K^+ \pi^-$ from the favored $b \to c$, 
while the $\bar{D^0}$ meson from the favored $b \to u$ suppressed amplitude 
is reconstructed in the favored decay $\bar{D^0} \to K^+ \pi^-$.     
By ignoring the possible effect due to the $D$ meson mixing, we define the the charge ratio of
$B^+$ and $B^-$ decay rate to the ADS final states $R^+$ and $R^-$, respectively.
\begin{equation}
R^+ = \frac{\Gamma(B^+ \to [K^- \pi^+] K^+)}{\Gamma(B^+ \to [K^+\pi^-]K^+)} 
= r^2_B + r^2_D + 2 r_B r_D k_D~ cos(\gamma + \delta_B + \delta_D)
\label{eq-R+}
\end{equation}
and
\begin{equation}
R^- = \frac{\Gamma(B^- \to [K^+ \pi^-] K^-)}{\Gamma(B^- \to [K^-\pi^+]K^-)}
= r^2_B + r^2_D + 2 r_B r_D k_D~ cos(\gamma - \delta_B + \delta_D)
\label{eq-R-}
\end{equation}
where
$r_B$ and $r_D$ are the suppressed to favored $B$ and $D$ amplitude ratios as
\begin{equation}
r_B \equiv \frac{|A(B^+ \to D^0 K^+)|}{|A(B^+ \to \bar{D^0} K^+))|}
\label{eq-rB}
\end{equation}
and
\begin{equation}
r^2_D = \frac{\Gamma(D^0 \to K^+ \pi- )}{\Gamma(D^0 \to K^- \pi^+)}
\label{eq-rD}
\end{equation}
and 
$\delta_B$ and $\delta_D$ are the strong phase differences between the two $B$ 
and the two $D$, respectively.

\section{\babar\ Detector}

The results presented in this paper are based on the entire $B\bar{B}$ data sample collected with 
the \babar\ detector at the PEP-II asymmetric-energy $B$ factory at the SLAC National 
Accelerator Laboratory. The $B\bar{B}$ pairs are produced from the decays of 
$\Upsilon(4S)$ resonance (on-resonance) that originate in collisions of 9.0 $GeV$ electrons
and 3.1 $GeV$ positrons. The on-resonance data sample has a mean energy of 10.58 $GeV$ and
an energy rms spread of 4.6 $MeV$. The off-resonance (continuum) data sample has a center-of-mass 
(CM) energy 40 $MeV$ below the resonance.

A detailed description of the \babar\ detector and the algorithms used
for track reconstruction and particle identification is provided
elsewhere~\cite{babar_nim}. A brief summary is given here.
High-momentum particles are reconstructed by
matching hits in the silicon vertex tracker (SVT) with track elements
in the drift chamber (DCH). Lower momentum tracks, which do not leave
signals on many wires in the DCH due to the bending induced by a
magnetic field, are reconstructed in the SVT alone.
Electrons are identified by the ratio of the track momentum to the
associated energy deposited in the calorimeter (EMC), the transverse
profile of the shower, the energy loss in the drift chamber, and
information from a Cherenkov detector (DIRC).  
The \babar\ detector Monte Carlo simulation is based on GEANT4~\cite{geant4}.
We use EVTGEN~\cite{evtgen} to model the kinematics of $B$ meson decays 
and use JETSET~\cite{jetset} to model off-resonance process $\epem\to\qqbar$ 
($q$ = $u$, $d$, $s$, or $c$ quark).

\section{Data Sample}

In the GLW method~\cite{GLW_babar}, we use $(467 \pm 5) \times 10^6$ \BB pairs, approximately 
equally divided into \BzBzb and \BpBm. The data have been collected in the years from 1999
until early 2008. We have reconstructed $B^{\pm} \to DK^{\pm}$ decays, 
with $D$ mesons decaying to non-$CP$ ($K\pi$), and $CP$-even $(K^+K^+, \pi^+\pi-)$, 
and $CP$-odd $(K_s^0 \pi^0, K_s^0\phi, K_s^0\omega)$ eigenstates.

In the ADS method, we use two results where $D^0$ mesons are reconstructed into two modes,
$D^{0}\to K^+\pi^-$~\cite{ADS1_babar} and $D^{0}\to K^+\pi^-\pi^0$~\cite{ADS2_babar}, respectively.
In $D^{0}\to K^+\pi^-$ mode we use a data of $(467 \pm 5) \times 10^6$ \BB pairs. 
We present a search of the decays $B^- \to D^{(*)} K^-$, where the neutral $D$ mesons decay 
into $K^+\pi^-$ final state (WS). In this paper, we first applied to 
$B^- \to D^{(*)} \pi^-$, where the neutral $D$ mesons decay into the Cabibbo-favored 
$(K^- \pi^-)$ and doubly suppressed mode. In $D^{0}\to K^+\pi^-\pi^0$, we use 
$(474 \pm 5) \times 10^6$ \BB pairs. An additional off-resonance data sample of 45 $fb^{-1}$ ,
collected at a center-of-mass energy 40 $MeV$ below the $\Upsilon(4S)$ resonance, is used
to study the $\epem\to\qqbar$ background. In this paper, we studied the decays $D^0$ and $\bar{D^0}$ 
in which decays to $K^{\mp}\pi^{\pm}\pi^0$ final states.

\section{Measurement of the CKM angle \boldmath $\gamma$}

\subsection{CP Observables in \boldmath $B^{\pm} \to D_{CP}K^{\pm}$ (GLW)}

We identify signal $B \to DK$ and $B \to D\pi$ candidates using two defined kinematic variables.
The first variable is the difference between the CM energy of the $B$ meson $(E^*_B)$ and 
the beam energy $(\Delta E)$ and the beam-energy-substituted mass $(m_{ES})$, respectively.
\begin{equation}
\Delta E = E^*_B - \sqrt{s}/2
\end{equation}
We estimate the irreducible background yields in our sample by exploiting the fact that 
the $D$ invariant mass distribution fir this background is approximately uniform, while
the signal event peaks around the nominal $D$ mass.

\begin{figure}[h]
\begin{center}
\begin{tabular}{lr}\hspace{-4mm}
\includegraphics[width=8.2cm]{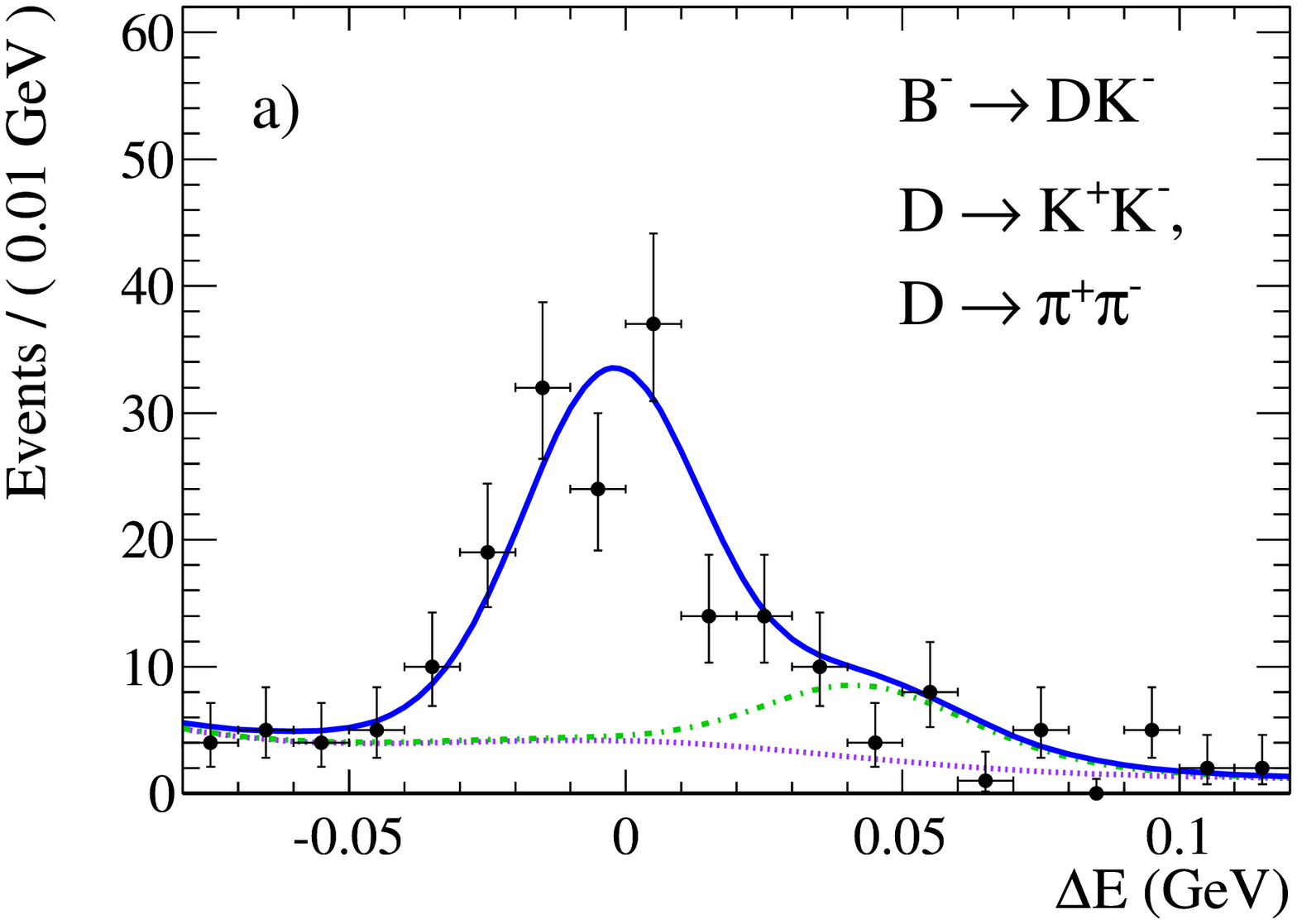}& \hspace{-1mm}  
\includegraphics[width=8.2cm]{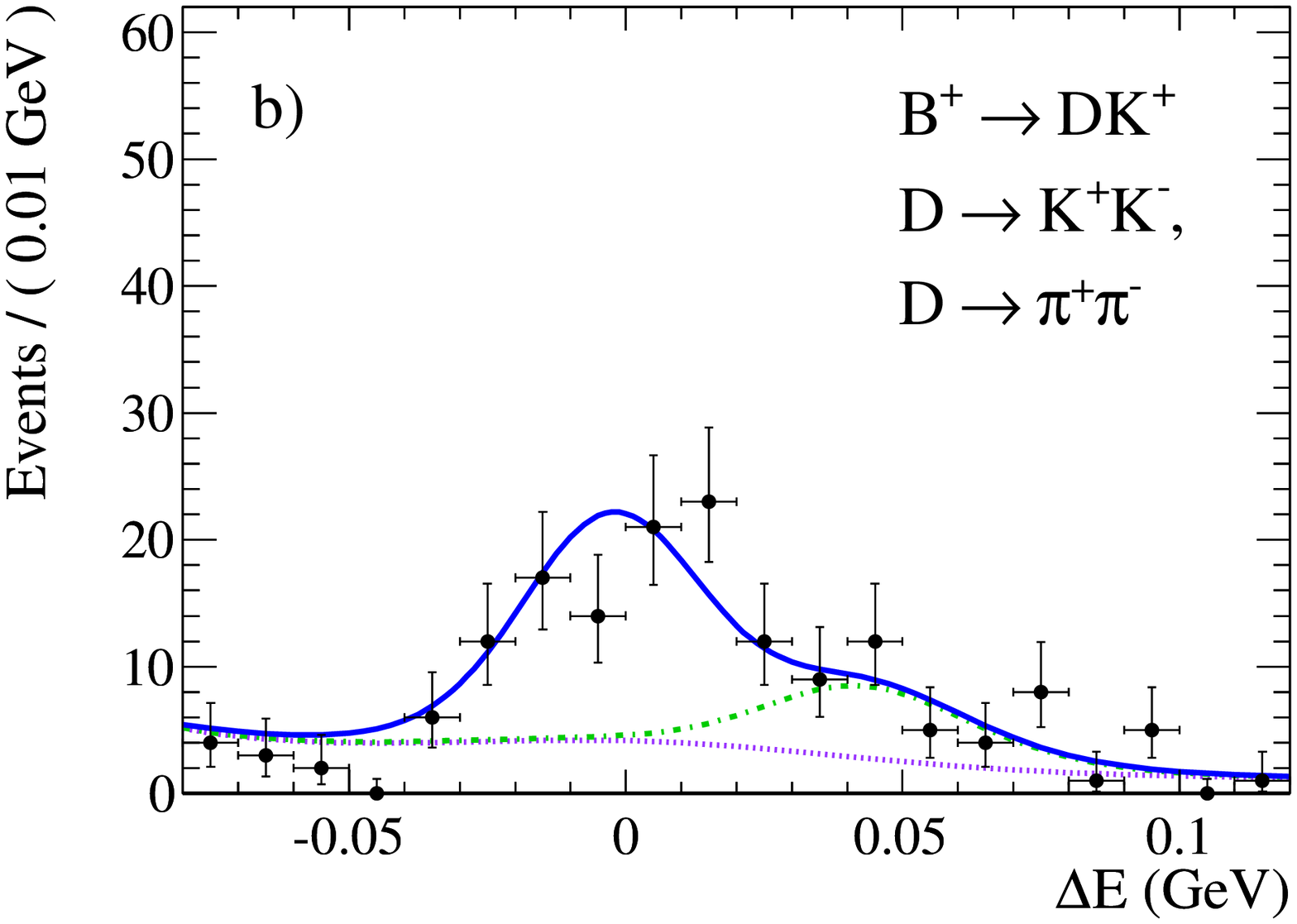} 
\end{tabular}
\caption{$\Delta E$ projections of the fits to the data, split into subset of
define $CP$ of the $D$ candidate and charge of the $B$ candidate: 
a) $B^- \to D_{CP+}K^-$ and b) $B^+ \to D_{CP+}K^+$} 
\label{GLW_deltaE}
\end{center}
\end{figure}

\begin{figure}[h]
\begin{center}
\begin{tabular}{lr}\hspace{-4mm}
\includegraphics[width=8.2cm]{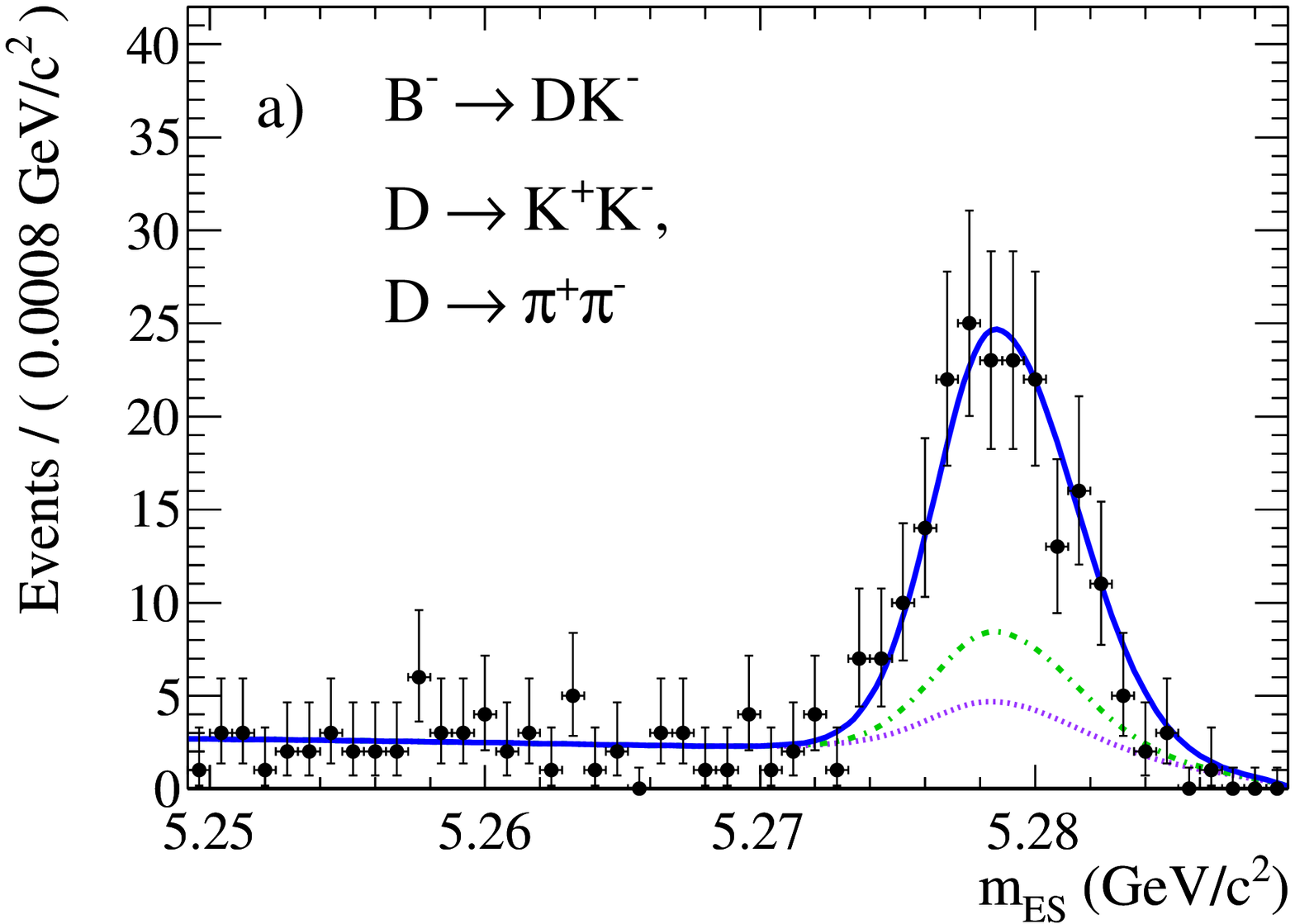}& \hspace{-1mm}  
\includegraphics[width=8.2cm]{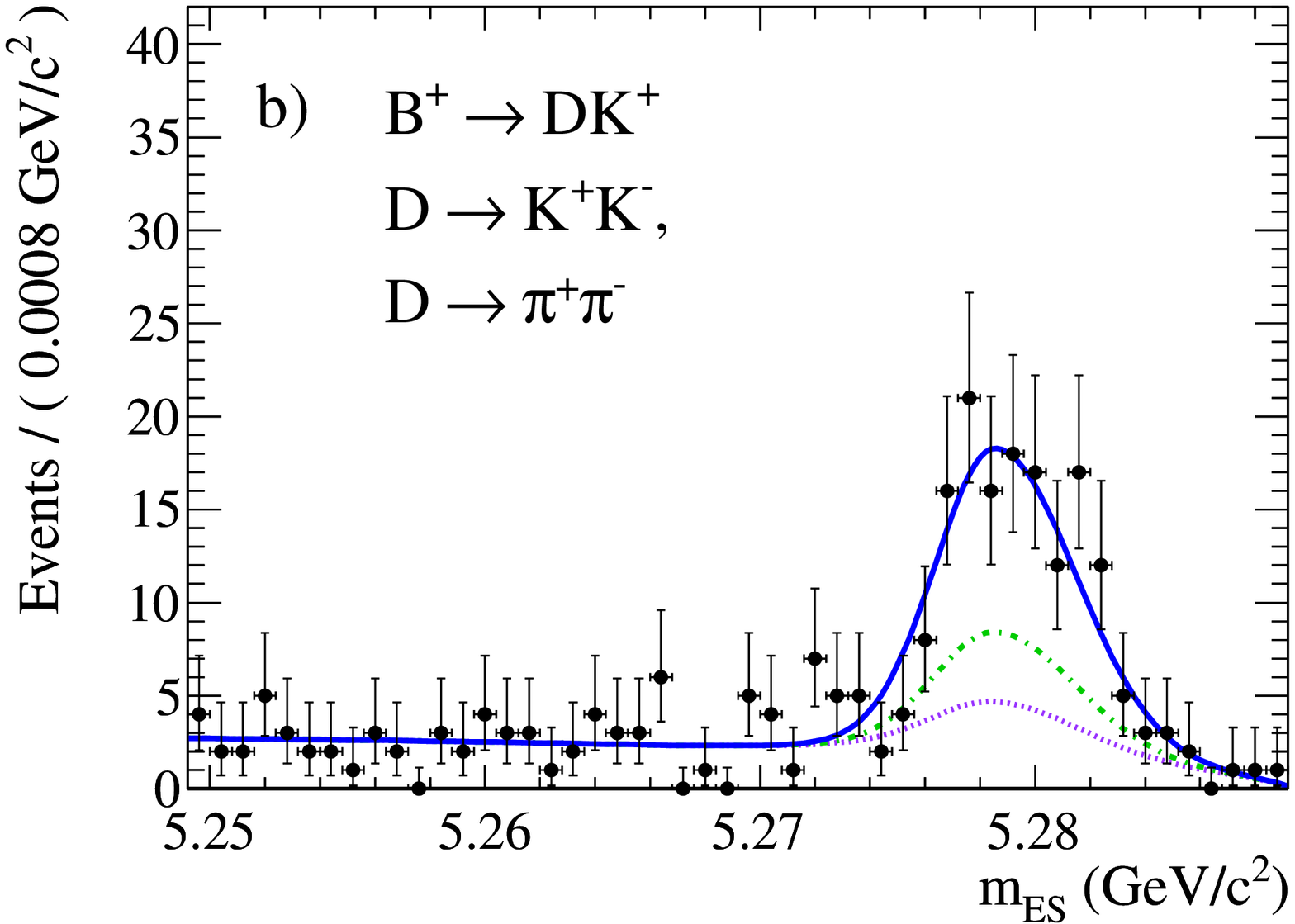} 
\end{tabular}
\caption{$m_{ES}$ projections of the fits to the data, split into subset of
define $CP$ of the $D$ candidate and charge of the $B$ candidate:
a) $B^- \to D_{CP+}K^-$ and b) $B^+ \to D_{CP+}K^+$} 
\label{GLW_mes}
\end{center}
\end{figure}

Figure~\ref{GLW_deltaE} shows the $\Delta E$ projections of the 
final fits to the $CP$ subsamples. The curves are the full PDF (solid, blue) 
and $B \to D\pi$ (dash-dotted, green) stacked on the remaining backgrounds 
(dotted, purple). The region between the solid and the dash-dotted lines represents 
the contribution of $B \to DK$. Figure~\ref{GLW_mes} shows $m_{ES}$ 
projections as well as projections to the fit to the $D^0 \to K^- \pi^+$ flavor mode.
The line definitions are the same as described in Fig.~\ref{GLW_deltaE}. 
We obtain the most precise measurements of the GLW parameters $A_{CP\pm}$ and $R_{CP\pm}$:
\begin{equation}
A_{CP+} = 0.25  \pm 0.06(stat) \pm 0.02(syst) \hspace{2cm}  A_{CP-} = -0.09 \pm 0.07(stat) \pm 0.02(syst)
\end{equation}
\begin{equation}
R_{CP+} = 1.18  \pm 0.09(stat) \pm 0.05(syst) \hspace{2cm}  R_{CP-} = 1.07  \pm 0.08(stat) \pm 0.04(syst)
\end{equation}
We measure a value of $A_{CP+}$ which is 3.6 standard deviations from zero, which constitutes 
the first evidence for direct $CP$ violation in $B \to DK$ decays.

\subsection{CP Observables in \boldmath $B^- \to DK^-$ and $B^- \to D^*K^-$ (ADS)}

We reconstruct $B^- \to D^{(*)}K^-$ and $B^- \to D^{(*)} \pi^-$ with the 
$D$ meson decaying to $K^-\pi^+$ (RS = right-sign) 
and $K^+\pi^-$ (WS = wrong-sign). Charged conjugate reactions are assumed 
throughout this paper. For decays involving a $D^*$, both $D^* \to D\pi^0$ 
and $D^* \to D\gamma$ are reconstructed. 
To study the \BB background for each signal category and charge combination
(RS and WS) we use a sample of $e^+e^- \to \Upsilon(4S) \to \BB$ Monte Carlo 
events corresponding to about 3 times the data luminosity. 

In this paper the off-resonance background events are reduced by using a multilayer 
perceptron artificial neural network with 2 hidden layers, available in the framework
of TMVA package~\cite{nn}. We use the neural network to select the discriminating 
variables.

\begin{figure}[h]
\begin{center}
\begin{tabular}{lr}\hspace{-4mm}
\includegraphics[width=8.6cm]{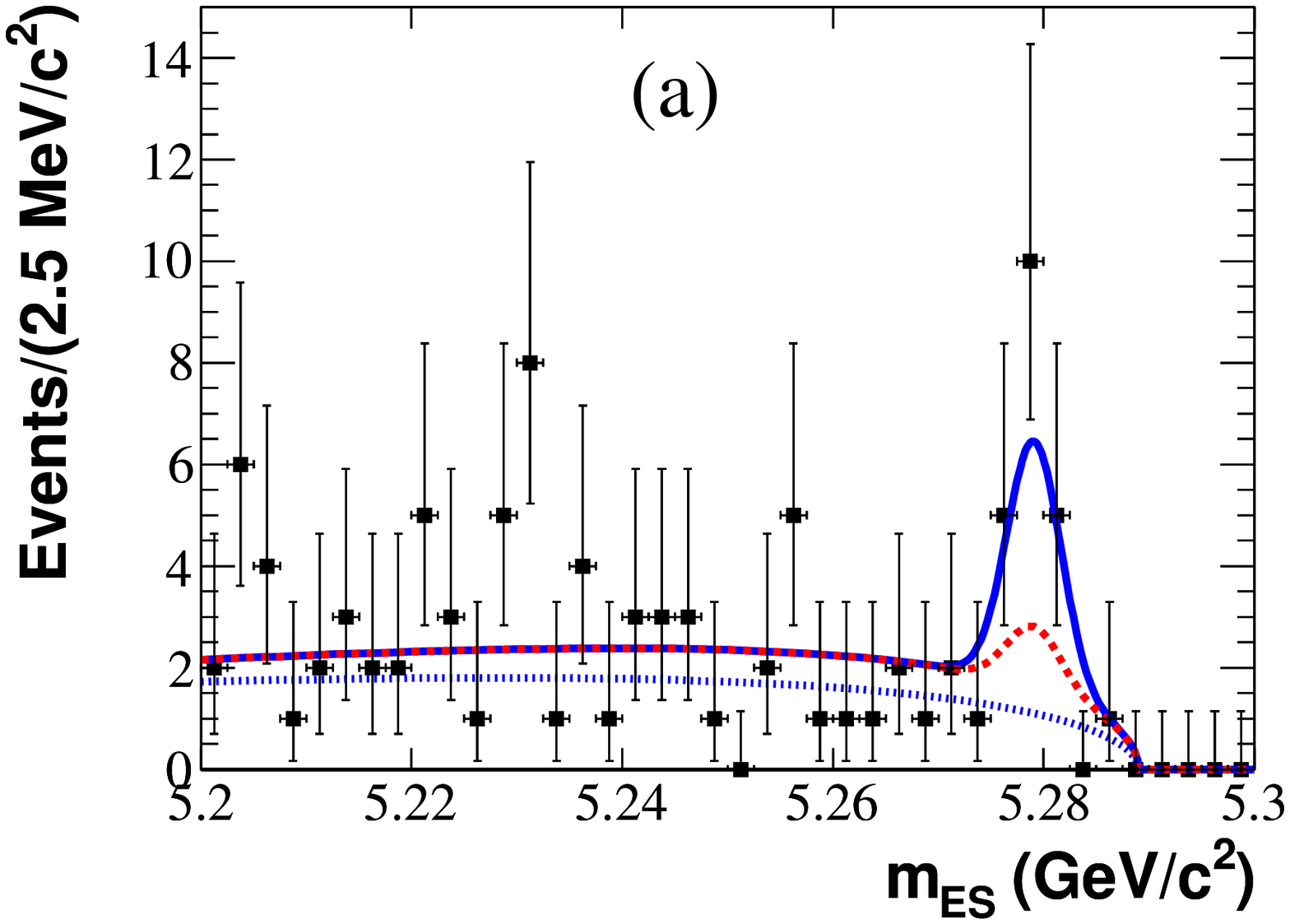}& \hspace{-6mm}  
\includegraphics[width=8.6cm]{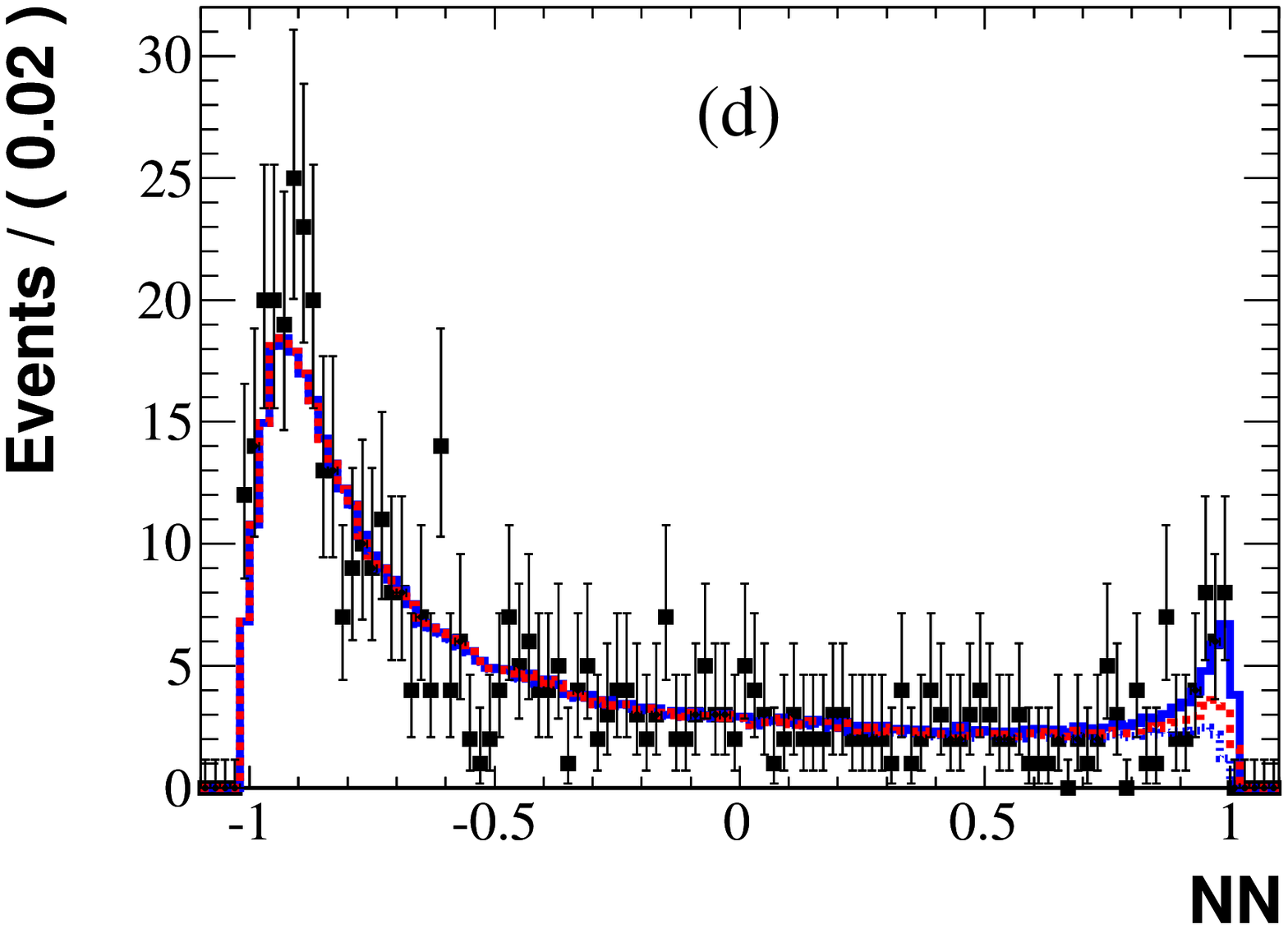} 
\end{tabular}
\caption{Projections on $m_{ES}$ and neural network 
(NN) of the fit results for $DK^+$ mode for samples}
\label{ADS_mes_nn}
\end{center}
\end{figure}

Figure~\ref{ADS_mes_nn} shows the projections on $m_{ES}$ and neural network 
(NN) of the fit results for $DK^+$ mode for samples enriched in signal with
the requirements $NN > 0.94$ for $m_{ES}$ projections or 
$5.2725 < m_{ES} < 5.2875$ $GeV/c^2$. 
The point with error bars are data. The curves represent the fit projections for 
signal plus background (solid), the sum of all background components (dashed), 
and \qqbar background only (dotted). The results of fits to the $B^+$ 
and $B^-$ sample
\begin{equation}
R^+ = (2.2 \pm 0.9 \pm 0.3) \times 10^{-2} \hspace{2cm} R^- = (0.2 \pm 0.6 \pm 0.2) \times 10^{-2}
\end{equation}
and we extracted the variables $r^{(*)}_B$
\begin{equation}
r_B = (9.5^{+5.1}_{-4.1})\%  \hspace{2cm} r^*_B = (9.6^{+3.5}_{-5.1})\%
\end{equation} 
\begin{figure}[h]
\begin{center}
\begin{tabular}{lr}\hspace{-4mm}
\includegraphics[width=8.6cm]{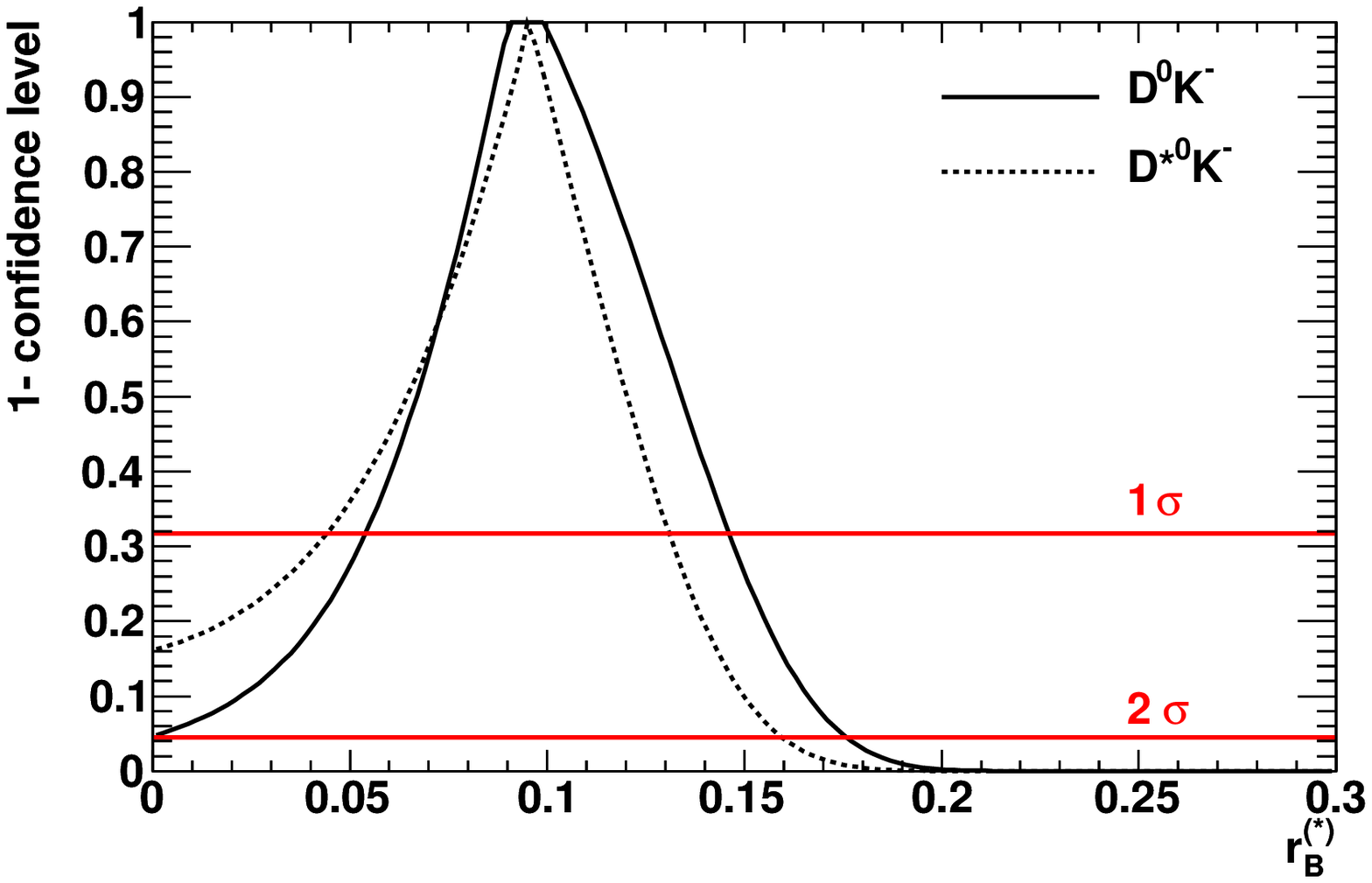}& \hspace{-6mm}  
\includegraphics[width=8.6cm]{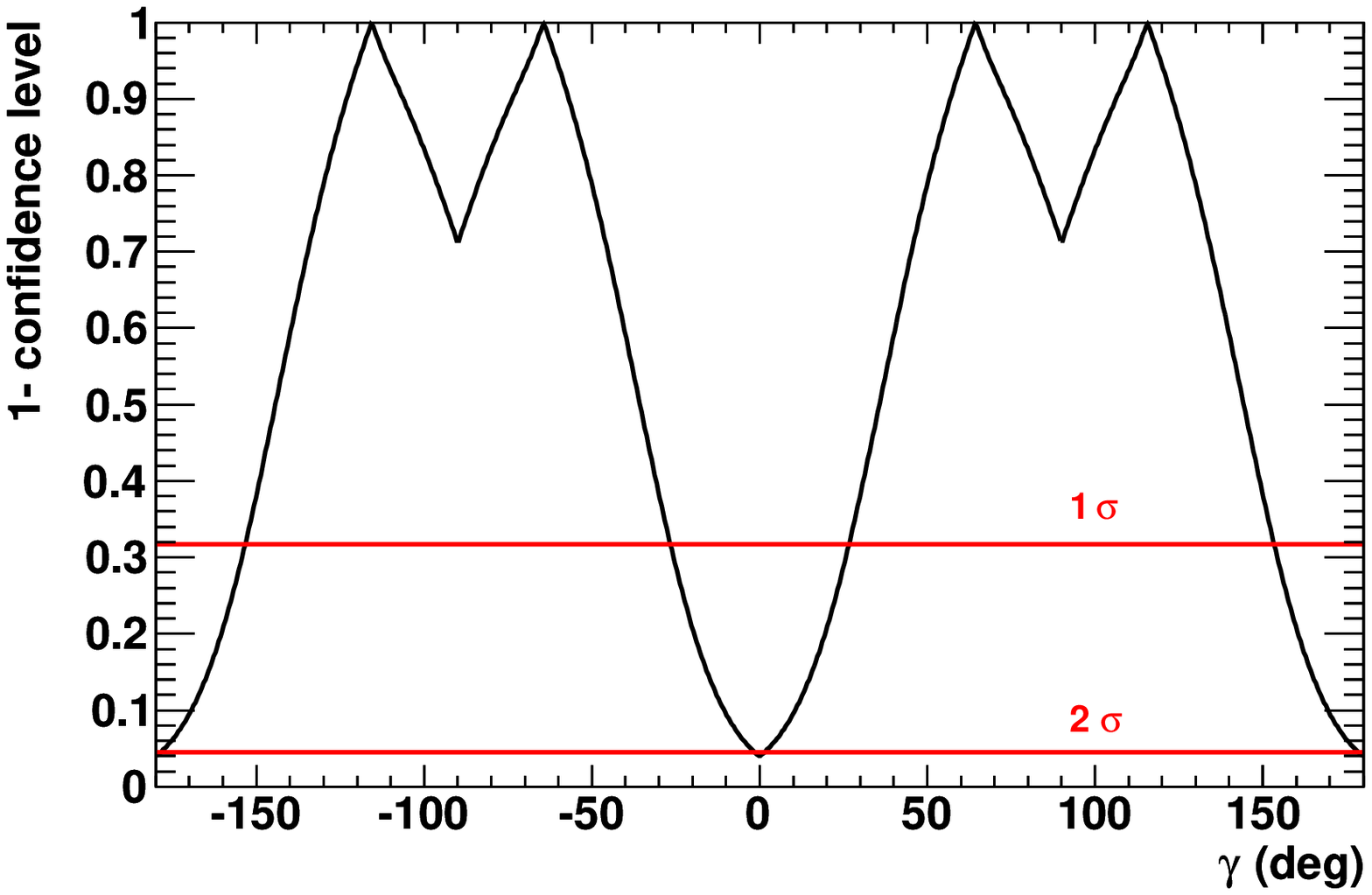} 
\end{tabular}
\caption{Constraints on $r^{(*)}_B$ from combined 
$B^- \to [K\pi]K^-$ ADS measurements (left) and constraints on
angle $\gamma$ from combined $B^- \to D^{(*)}[K^+ \pi^-]K^-$ 
ADS measurements (right).}
\label{ADS_gamma}
\end{center}
\end{figure}

\subsection{CP Observables in \boldmath $B^{\pm} \to [K^{\mp} \pi^{\pm}\pi^0]K^{\pm}$ (ADS)}

In this paper the off-resonance background events, in contrast to \BB events, are 
characterized by a jet-like topology. We use a Fisher discriminant $\cal{F}$ to discriminate 
between the two categories of events. The Fisher discriminant is a linear combination
of six variables. We choose the coefficients of the linear combination to maximize the
separation between the signal and the off-resonance background. For the signal the 
Fisher discriminant $\cal{F}$ peaks at 1 and -1 for the off-resonance events.
Since the correlations among the variables are negligible, we write the PDFs as products
of one dimensional distributions of the $m_{ES}$ and $\cal{F}$. We use MC samples to check
the absence of the correlation between these distributions.

The PDF parameters are derived from data sample when possible in order to reduce 
the systematic uncertainties. The parameters for the continuum events are determined
from the off-resonance data sample. We use the data sample of $B^+ \to D\pi^+$ 
with $D \to K^+ \pi^- \pi^0$ to extract the parameters for the $m_{ES}$ distribution.
The parameters for the non-peaking \BB distributions and the signal Fisher PDF are 
determined from the MC sample.

\begin{figure}[h]
\begin{center}
\begin{tabular}{lr}\hspace{-4mm}
\includegraphics[width=8.8cm]{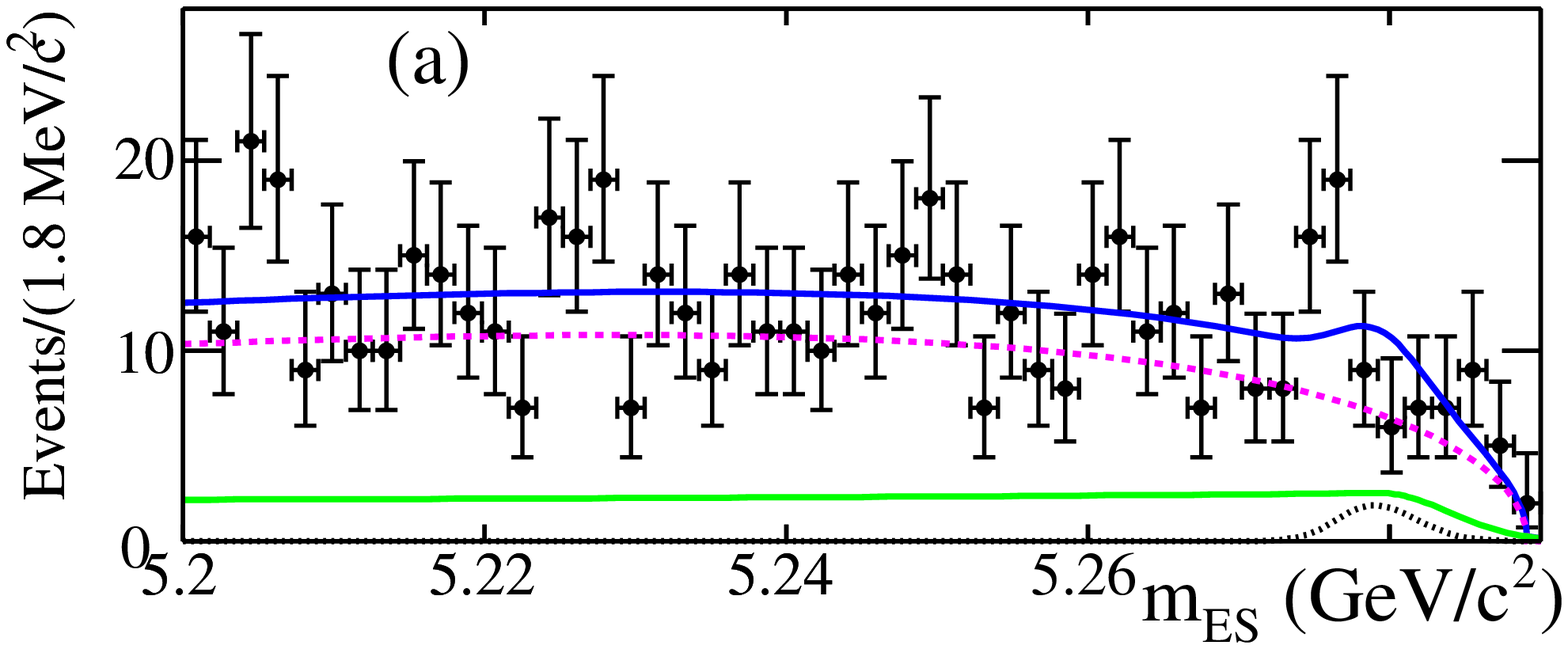}& \hspace{-6mm}  
\includegraphics[width=8.8cm]{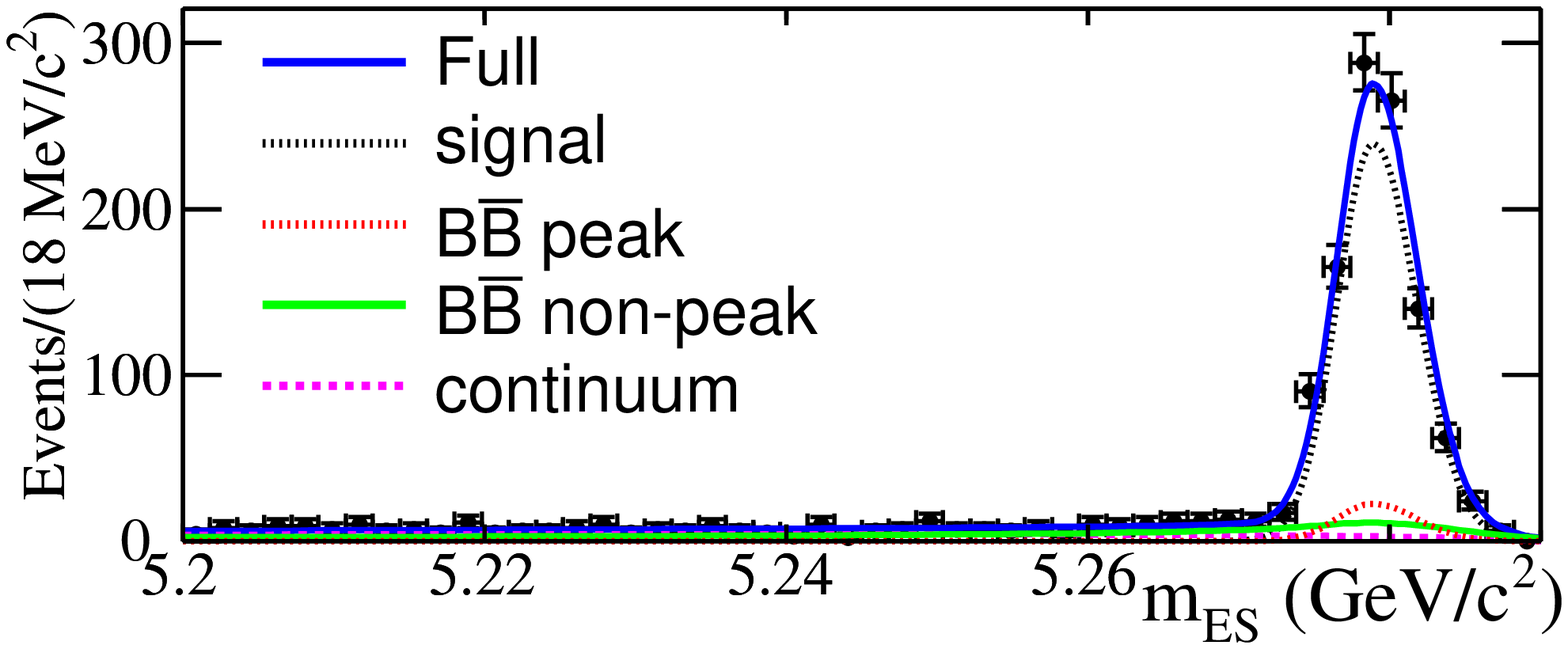} 
\end{tabular}
\caption{Distribution of $m_{ES}$ (a,b) with $\cal{F} >$ 0.5.}
\label{ADS_new1}
\end{center}
\end{figure}

\begin{figure}[h]
\begin{center}
\begin{tabular}{lr}\hspace{-4mm}
\includegraphics[width=8.8cm]{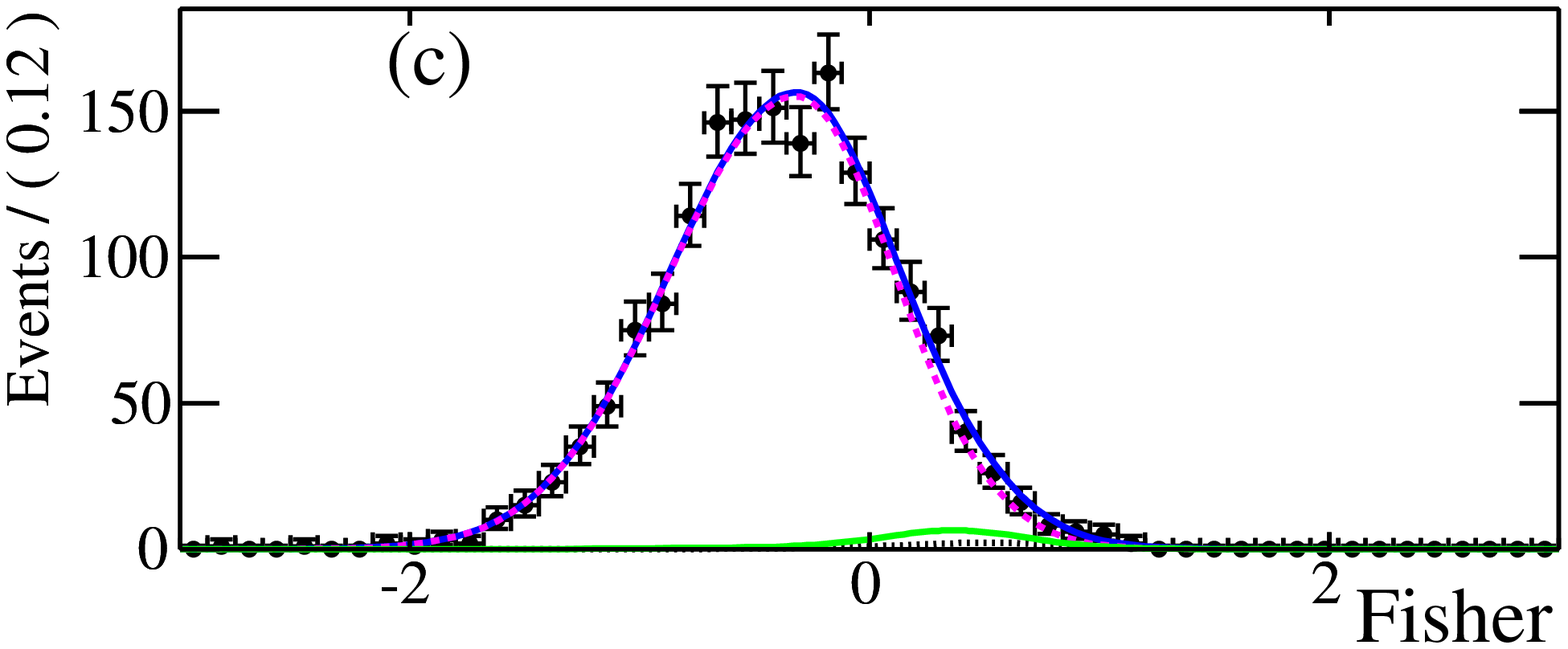}& \hspace{-6mm}  
\includegraphics[width=8.8cm]{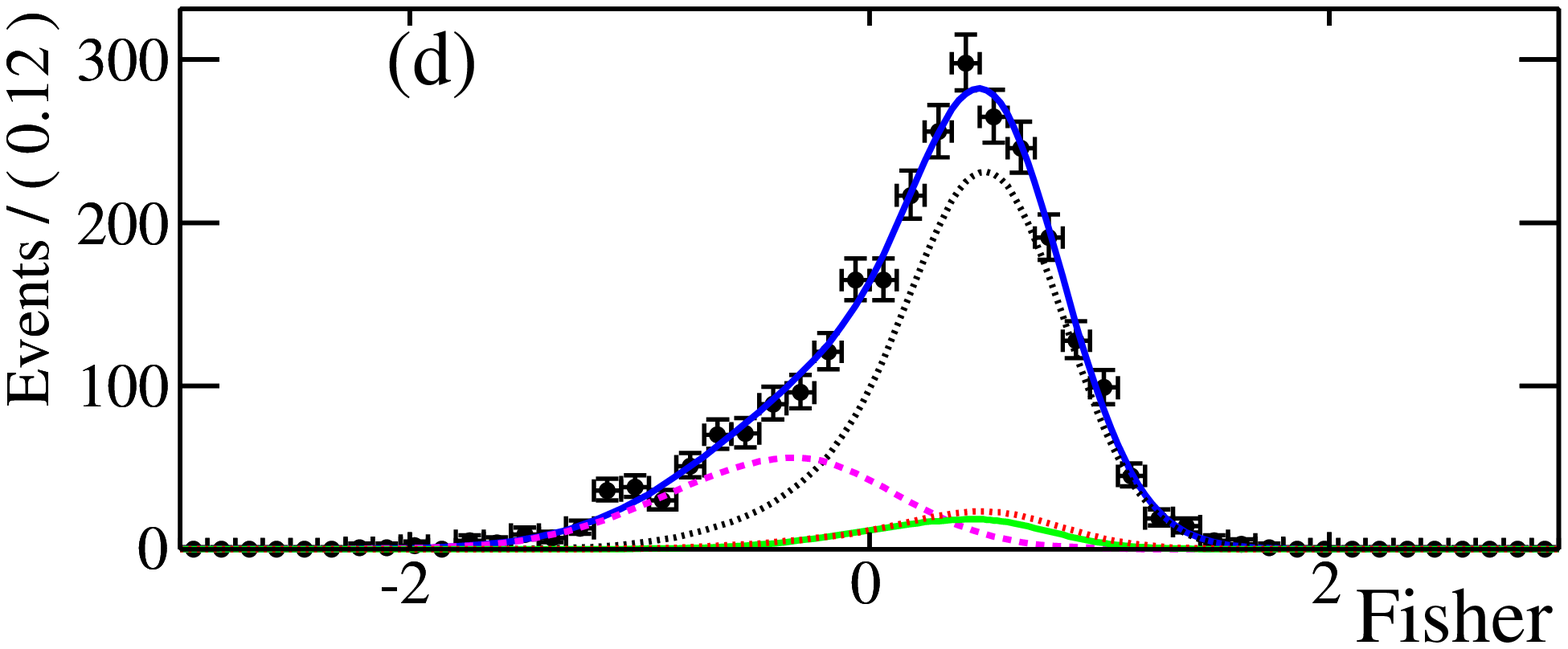} 
\end{tabular}
\caption{Distribution of $\cal{F}$ (c,d) with $m_{ES} > 5.27$ GeV/$c^2$.}
\label{ADS_new2}
\end{center}
\end{figure}

The fits to the $m_{ES}$ for $\cal{F} >$ 0.5 and the Fisher discriminant $\cal{F}$
distribution with $m_{ES} > 5.27$ GeV/$c^2$ for the combined $B^+$ and $B^-$ samples 
are shown in Fig.~\ref{ADS_new1} and Fig.~\ref{ADS_new2}, respectively. 
The data are well described by the overall fit results (solid blue line) which is
the sum of the signal, continuum background, non-peaking \BB background, 
and peaking \BB background.
We have presented a study of the decays $B^{\pm} \to D^0 K^{\pm}$ 
and $B^{\pm} \to \bar{D^0} K^{\pm}$, in which the $D^0$ and $\bar{D^0}$ mesons decay
to $K^{\pm}\pi^{\pm}\pi^0$ final state using the ADS method. The final results are
\begin{equation}
R^+ = (5^{+12+1}_{-10-4})  \times 10^{-3} \hspace{2cm} R^- = (12^{+12+2}_{-10-4}) \times 10^{-3}
\end{equation}
from which we obtain 90\% probability limits
\begin{equation}
R^+ < 23 \times 10^{-3} \hspace{2cm} R^- < 29 \times 10^{-3}
\end{equation}
Following a Bayesian approach~\cite{bayesian1,bayesian2}, the probability distributions 
for the $R^+$ and $R^-$ ratio obtained in the fit are translated into a probability distribution 
for $r_B$.  Figure~\ref{ADS_gamma} shows the posterior probability distribution. 
\begin{figure}[h]
\begin{center}
\includegraphics[width=8.6cm]{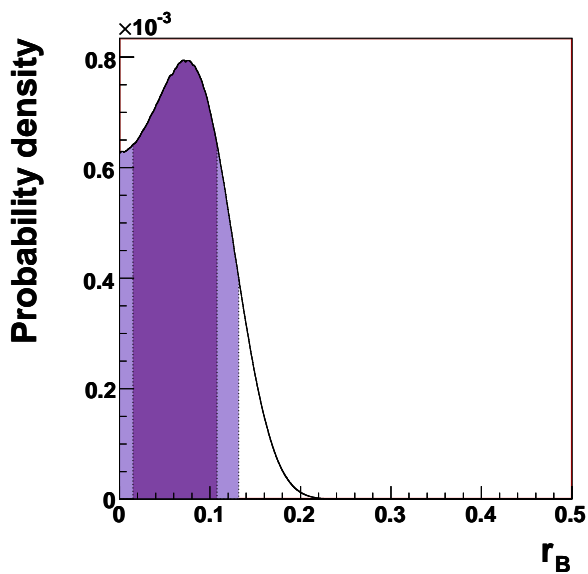} 
\caption{Bayesian posterior probability density function for $r_B$ from our measurement
of $R^+$ and $R^-$ and the hadronic $D$ decay parameters $r_D$, $\delta_D$, and $k_D$.}
\label{ADS_gamma}
\end{center}
\end{figure}
Since the measurements are not statistically significant, we integrate over the positive 
portion of that distribution and obtain the upper limit $r_B < 0.13$ at 90\% probability, 
and the range 
\begin{equation}
r_B \in [0.01, 0.11] 
\end{equation}
at 68\% probability and 0.078 as the most probable value.

\begin{acknowledgments}
The author would like to thank the organizers of DPF 2011 for their excellent program 
and kind hospitality. The supports from the \babar\ Collaboration, the University of 
South Alabama, and the University of Mississippi are gratefully acknowledged. 
This work was supported by the U.S. Department of Energy under 
grant No. DE-FG02-96ER-40970.
\end{acknowledgments}

\bigskip 

\end{document}